\documentstyle[12pt]{article}
\begin{document}
\newcommand{\beq}{\begin{equation}}
\newcommand{\eeq}{\end{equation}}

\begin{flushright}
ITEP--TH--41/00\\
hep-th/0007247
\end{flushright}
\vspace{0.5cm}
\begin{center}
{\Large\bf On noncommutative vacua and \\[2mm]noncommutative solitons}\\
\end{center}
\bigskip
\begin{center}
{\bf A.S.Gorsky, Y.M.Makeenko and K.G.Selivanov}
\bigskip

{ ITEP,  B.Cheremushkinskaya 25, 117259 Moscow}
\end{center}
\bigskip
\begin{abstract}
We consider noncommutative
theory of a compact scalar field.
The recently discovered projector solitons are interpreted as
classical vacua in the model considered.
Localized solutions to the projector equation are pointed out and their
brane interpretation is discussed. An example of the
noncommutative soliton interpolating between such vacua is given.
No strong noncommutativity limit is assumed.
\end{abstract}

1.   Noncommutative theories have been
attracted recently a great deal of attention
both at perturbative and nonperturbative levels. Even the
simplest noncommutative scalar theory reveals a new peculiar object
which is absent in commutative theory namely nontrivial solution
to the equation of motion at large noncommutativity.
It was identified as the noncommutative soliton in the scalar
theory with generic potential~\cite{gms,mukhi}. Later this solution
got interpreted as a brane localized within the brane of higher
dimension with the large B field included~\cite{harvey}. This solution
finds a proper place in the context of the tachyon condensation
\cite{sen} as well as in the open string field theory formalism
\cite{witten}. Recently new solutions implying the restoration
of the full gauge symmetry  outside the soliton core have been
found~\cite{gms2}.

In this note we would like to consider the theory with a compact
scalar field. In this case we shall show that noncommutativity
amounts to  rather rich vacuum structure. Interestingly the
projector solitons discussed above,
despite the fact that they have a nontrivial dependence
on noncommutative coordinates,
play in our case the role of classical vacua
additional to the simple ``commutative'' ones.
A domain wall type soliton interpolating between
the vacua is found. We also construct a new localized
solution in the case of large noncommutativity. The brane
interpretation of the solutions considered is given.

2. Let us remind the description of the noncommutative solitons
in terms of the projector operators.
We shall work in (2+1)D space with Euclidean signature, two coordinates
being noncommutative:  \beq \label{commutator} [x,y]=-i.  \eeq In this
equation we assume that the noncommutative coordinates have been rescaled to
get rid of the noncommutativity parameter $\theta$ from the right hand side;
it will then be present in front of a kinetic term (cf.~\cite{gms}).
In~\cite{gms} solitons in a theory of noncommutative scalar field
were obtained in the limit
of strong noncommutativity ($\theta \rightarrow \infty$).
In that limit one can neglect kinetic term and the field equation reduces to
\beq
\label{fe}
\frac{\partial}{\partial \phi} V(\phi)=0,
\eeq
where $V(\phi)$ is a potential, $V(\phi)=1/2m^{2} \phi^{2} +
\sum_{n} 1/n!\lambda_{n} \phi^{n} $.
The solutions to Eq.(\ref{fe}) were found using the ansatz
\beq
\label{anzatz}
\phi(x,y)=\sum_{n} \alpha_{n} P_{n}(x,y)
\eeq
where $P_{n}(x,y)$ is a set of projectors,
\beq
\label{pro}
P_{n}(x,y) \ast P_{m}(x,y) = \delta_{n,m} P_{n}(x,y),
\eeq
and the star-product $f(x^{\mu}) \ast g(x^{\nu}) =
e^{-i/2 \epsilon_{\mu \nu} \frac{\partial}{\partial_{\eta_{\mu}}}
\frac{\partial}{\partial_{\rho_{\nu}}}}f(\eta_{\mu})
g(\rho_{\nu})|_{\eta_{\mu}=x^{\mu}, \rho_{\nu}=x^{\nu}}$
have been used.

In view of Eq.(\ref{pro}) the field equation~(\ref{fe}) further reduces
to
\beq
\frac{\partial}{\partial \phi} V(\alpha_{n})=0
\eeq
so the parameters $\alpha_{n}$ must be extrema of the potential.
This way the problem of constructing solutions has been reduced to the problem
of constructing projectors.

Radially symmetric projectors, $P_{n}(r^{2})$,
are most conveniently  described (see \cite{mukhi})
in terms of a generating function
$P(r^{2}, u)$,
\beq
\label{gen}
P(r^{2}, u)=\sum_{n} u^{n} P_{n}(r^{2}).
\eeq
Eq.(\ref{pro}) is then rewritten as
\beq
\label{pro2}
P(r^{2}, u) \ast P(r^{2}, v) = P(r^{2}, uv)
\eeq
and the completeness condition
\beq
\label{comp}
\sum_{n} P_{n}=1
\eeq
is rewritten as
\beq
\label{comp2}
P(r^{2}, 1)=1.
\eeq
The solution to Eqs.(\ref{pro2}), (\ref{comp2}) reads
\beq
\label{gen2}
P(r^{2}, u)=\frac{2}{u+1}e^{\frac{u-1}{u+1} r^{2}}
\eeq
which is a generating functions for the Laguerre polynomials.
The construction of projectors completes the construction of the
noncommutative solitons~\cite{gms}.

Our point here is to interpret  those solutions rather as
classical vacuum states, while the proper solitons are solutions
interpolating between these vacua.  To illustrate this point, let us
consider the following noncommutative field theory
with the compact scalar field:
\beq
\label{lagrangian}
L=\int \frac{1}{2\theta}\partial_{A}g^{-1}\partial_{A}g + 2 - g - g^{-1}
\eeq
where $g=e_{\ast}^{i \phi}=1+ i\phi + 1/2 i\phi \ast i\phi + \ldots $
and $g^{-1}$ is inverse to $g$ in the sense of the star-product:
\beq
g^{-1}\ast g = g \ast g^{-1}=1.
\eeq
The
derivatives $\partial_{A}$ include derivatives with respect to
the noncommutative coordinates $\partial_{\mu}$ and derivative with
respect to (Euclidean or Minkowski) time $\tau$.

Notice that the kinetic term in Eq.(\ref{lagrangian}) can be considered as
the one of a noncommutative $U(1)$ sigma-model. In the commutative limit
it reduces
to the common kinetic term for a scalar field so it is a legal noncommutative
generalization of the latter. The potential term reduces in the commutative
limit to $2(1-\cos \phi)$.
In principle one can consider
additional vacua in the theory without the potential term
but it will be used to construct time-dependent solutions
interpolating between different vacua.
By the way, in 2D one can add
a Wess-Zumino term to the Lagrangian~(\ref{lagrangian}) obtaining this
way a noncommutative integrable sin-Gordon model. At rational values of
the noncommutativity parameter the noncommutative sin-Gordon model
on a torus is
Morita equivalent to a nonabelian sin-Gordon model, a member of the family
of Polyakov's nonabelian Toda theories.  The noncommutative sin-Gordon
model was discussed in the context of the noncommutative
bosonization in \cite{nunez}.

Let us return back to the model Eq.(\ref{lagrangian}). Corresponding field equation
reads
\beq
\label{fe2}
\frac{1}{\theta}\partial_{A} \left (g^{-1} \ast \partial_{A}g \right )
= g - g^{-1}.
\eeq
In complete analogy with the commutative case there are classical vacua
solutions
\beq
\phi_{n}=2\pi n.
\eeq
In addition there are solutions of the type of Eq.(\ref{anzatz}) with
\beq
\label{vacua}
\alpha_{n}=2\pi m_{n}
\eeq
where $m_{n}$ is a set of integers
(if one considers a theory without the potential, $\alpha_{n}$
are no longer restricted to be integer).
These are obviously solutions of
Eq.(\ref{fe2}) in view of the fact that when $\phi$ is of the form of
Eq.(\ref{anzatz}),
\beq
\label{g}
g=e_{\ast}^{i\phi}=\sum_{n} e^{i\alpha_{n}} P_{n}
\eeq
where, we remind, $e_{\ast}^{i\phi}$ is the star-exponential of $i\phi$
and the exponential of $\alpha$ on the right hand side
is the usual exponential.
When $\alpha_{n}$ are as in Eq.(\ref{vacua}), $g$ is equal to 1,
so such states are most natural to interpret as the additional classical vacua.
We would like to emphasize that unlike~\cite{gms} we do not restrict ourselves
to the strong noncommutativity limit and vacuum states exist at arbitrary
$\theta$.

Let us discuss in more detail the vacuum state of the type
\begin{eqnarray}
\label{delta}
\alpha_{n}&=&\alpha_{+},\; n-{\rm even}\,, \nonumber\\
\alpha_{n}&=&\alpha_{-},\; n-{\rm odd}\,.
\end{eqnarray}
Then
\beq
\phi=\alpha_{+} \frac{1+P(r^{2}, -1)}{2} +
\alpha_{-} \frac{1-P(r^{2}, -1)}{2}.
\eeq
Using the fact that
\beq
P(r^{2}, -1)=\pi \delta^{2}(x)
\eeq
(notice that, in view of Eq.(\ref{pro2}),
$\pi \delta^{2}(x) \ast \pi \delta^{2}(x)=1$)
one can also rewrite the field as
\beq
\phi=\frac{\alpha_{+}+\alpha_{-}}{2} +
\frac{\alpha_{+}-\alpha_{-}}{2} \pi \delta^{2}(x).
\eeq
According to the general discussion above, it is a classical
vacuum state, providing $\alpha_{+/-}=2\pi m_{+/-}$.
A few additional remarks concerning this solution are in order. First,
the soliton can be placed at an arbitrary point
so a nontrivial
moduli space of the solutions appears. We shall describe it in a moment.
One could ask if a superposition
of the localized solitons placed at different positions in the
noncommutative plane is again a solution to the equation
of motion. The answer is ``no'' due to the following identity
\beq
\pi \delta^{2}(x-a) \ast \pi \delta^{2}(x-b)=e^{i\Phi(a,b,x)}
\eeq
where $\Phi(a,b,x)=2\epsilon_{\mu\nu}(a_\mu b_\nu+b_\mu x_\nu + x_\mu a_\nu)
$ is the flux of the B field through the triangle
with the vertices at $a$, $b$ and $x$.

The solution~(\ref{anzatz}) can be
transformed into another solution by a star-unitary
transformation
\begin{equation}
\phi^\prime
=\omega*\phi*\omega^{-1}.
\label{starunitary}
\end{equation}
This is again a solution since
$\omega*\omega^{-1}=1$.
The translation discussed above is
the simplest example of such a transformation given by
$\omega=e^{ik_x x+ik_y y}$.
Another example is given by
the Gaussian function
\begin{equation}
\omega (x,y)= \sqrt{1+\alpha\beta -\gamma^2}\;
e^{i\alpha x^2 +i \beta y^2 + 2i\gamma xy}
\end{equation}
which is star-unitary. The transformation generated by
this function keeps the functional form of $\phi(x,y)$ unchanged but
rotates $x,y$ in the noncommutative plane
by an $sl(2;R)$ matrix into
\begin{eqnarray}
x^\prime &=&
\frac{(1+\gamma^2)^2-\alpha\beta}{1+\alpha\beta -\gamma^2}\,x +
\frac{2\beta}{1+\alpha\beta -\gamma^2}\, y \,,\nonumber \\*
y^\prime &=& -\frac{2\alpha}{1+\alpha\beta -\gamma^2}\, x   +
\frac{(1-\gamma^2)^2-\alpha\beta}{1+\alpha\beta -\gamma^2}\,y \,.
\end{eqnarray}
In general, the new solution is no longer spherically symmetric.
Some of the star-unitary transformation leave the solution~(\ref{anzatz})
invariant. An example of such a star-unitary function is
\begin{equation}
\omega (x,y)= \pi \delta (x)\delta (y)
\end{equation}
which only inverts the signs
of $x$ and $y$.

3. Thus the structure of vacua in the noncommutative framework is much
more involved
than in the commutative one.
Let us now discuss solitons of the type of domain walls, i.e.,
solutions interpolating between different vacua of the type of
Eq.(\ref{anzatz}). We shall now assume that
coefficients $\alpha$ in Eq.(\ref{anzatz}) are $\tau$-dependent but
independent
of the noncommutative coordinates. In this assumption the field
equation~(\ref{fe2}) reduces to
\beq
\label{fe3}
i/\theta \sum_{n} (\frac{ \partial}{\partial{\tau}})^{2}\alpha_{n} P_{n}+
1/\theta \sum_{n,m} e^{i(\alpha_{m}-\alpha_{n})}
\partial_{\mu} \left ( P_{n} \ast \partial_{\mu} P_{m} \right )
=\sum_{n}(e^{i\alpha_{n}}-e^{-i\alpha_{n}})P_{n}
\eeq
The term with derivatives of the projectors,
$\partial_{\mu} \left ( P_{n} \ast \partial_{\mu} P_{m} \right )$,
can be computed in terms of the projectors using the generating function:
\begin{eqnarray}
\left (\partial_{\mu} P(r^{2}, u) \right ) \ast
\left (\partial_{\mu} P(r^{2}, v) \right )=
(u-1)(v-1)\partial_{uv} \left ((uv+1) P(r^{2}, uv) \right) \nonumber\\
\partial_{\mu} \left ( P(r^{2}, u) \ast \partial_{\mu} P(r^{2}, v) \right )=
(u+1)(v-1)\partial_{uv} \left ( (uv-1) P(r^{2}, uv) \right)
\end{eqnarray}
We  shall however not need it because we make a further assumption
under which that term disappears. Namely, we assume that
all $\alpha$'s differ from each other by only integers, that is,
\beq
\label{soliton}
\alpha_{n}=\bar{\alpha} + 2\pi m_{n}.
\eeq
Then Eq.(\ref{fe3}) reduces to an ordinary differential equation for
$\bar{\alpha}$,
\beq
(\frac{\partial}{\partial {\tau}})^{2}\bar{\alpha}=
2\theta \sin(\bar{\alpha}),
\eeq
the solitonic solution to which is  well known to be
\beq
\bar{\alpha}=\frac4{\sqrt{2\theta}} \arctan(e^{\sqrt{2\theta} \tau}).
\eeq
For this solution all $\alpha_{n}$'s change by $2\pi$ when $\tau$
runs from $-\infty$ to $\infty$.

One can introduce a topological charge associated to solitons,
a generalization
of $\int_{C} d \phi $ to the noncommutative framework:
\beq
\label{charge}
Q=\frac{1}{2\pi i}\int d^{2}x \int_{C} g^{-1} \ast dg
\eeq
where the first integration is over the noncommutative plane (a sort of trace)
and the contour $C$ is assumed to begin and end at the same point of the
noncommutative plane. Since the integrated form is a flat connection,
$Q$ does not depend on the shape of $C$
and does not change under
the star-unitary transformation~(\ref{starunitary}).
When $g$ is of the form Eq.(\ref{g})
with possibly $\tau$-dependent $\alpha$'s, one obtains
\beq
Q=\frac{1}{2\pi }\int d^{2}x \sum_{n} \alpha_{n}|^{+}_{-} P_{n}
=\sum_{n} \alpha_{n}|^{+}_{-}
\eeq
where $\alpha_{n}|^{+}_{-}$ is a total change of $\alpha_{n}$ along the path $C$.
The projectors are easily integrated using the generating function Eq.(\ref{gen2}).
One can also use non-integrated version of topological charge which depends on a point
on the noncommutative plane.

4. Let us discuss the brane interpretation of the solutions discussed
above. First let us consider the new solution to the projector equation
we have constructed. It was shown in~\cite{harvey} that noncommutative
soliton at large noncommutativity is nothing but a brane of lower
dimension inside the brane if the
tachyon Lagrangian is considered. For instance
the Gaussian projector solution for bosonic string
was interpreted as D23 brane within D25
brane. The key point of this identification is the reproducing of the
correct tension of D23 brane from the action calculated on the noncommutative
soliton. However this interpretation is a little bit subtle
since the size of the Gaussian soliton is nonzero.
For the localized delta function
solution which is perfectly localized this subtlety is resolved and to
confirm the identification of our localized solution as a D brane
we have to calculate its tension.

For example the action on the
localized soliton in bosonic string looks as
\beq
S= -\frac{g_sT_{25}}{G_s}\int d^{24} \int d^{2} \sqrt{G} \phi(r)
\eeq
where $\phi(r)$ is the localized soliton, $g_s$ is a string coupling,
G is a metric and $G_s= \frac{g_s \sqrt {G}}{ 2 \pi B \sqrt{g} \alpha^{'}}$.
The direct calculation amounts to the correct tension of the brane. Let us
also note that in the type II theory where the tachyon potential
is definitely even one can construct the additional solution
based on the relation $\Delta^3=\Delta$ instead of the
projector one solving the equation of motion.
This solution
\beq
\Delta(x)=\pi \delta^2(x)
\eeq
admits the similar brane interpretation as well.

Let us turn to the role of the solution in the theory of the compact
scalar which it presumably plays in the brane context. To this
aim remind that the complex tachyon field is certainly present
for the brane-antibrane system \cite{sen} and its potential has
a form of the ``Mexican hat''. Then at the vacuum manifold we have
$T \ast \bar{T} =const$ therefore the vacuum valley is
parameterized by the phase of the complex scalar which is definitely
compact. Therefore we  arrive at the theory with
the action we have discussed before. Hence we can expect nontrivial
vacuum states involving projectors in the theory of the complex tachyon.
It is known that the BPS D(p-2) brane arises as vortex in the complex
tachyon field in the $Dp-\bar{Dp}$ system \cite{sen} (noncommutative vortexes
were also discussed \cite{vortex})
Such stable
configuration can be also derived for the
additional noncommutative vacuum state where the
tachyon phase has nontrivial space dependence. We shall discuss this
point in more details elsewhere.

5.  In this letter we have described a rich vacuum structure in the
noncommutative theory of the compact scalar field. The example of the kink
type solution interpolating between two vacua of the theory is presented. New
solutions to the projector equations are found  and their brane
interpretation is given. Let us also note
that the noncommutative solitons where recently discussed in the context of
Quantum Hall effect \cite{pas}. It seems that the new localized
noncommutative solitons we have found are better candidates
for the skyrmion spin textures then Gaussian solitons discussed in that
paper.

This work was supported in part by CRDF grant RP1-2108.
The work of A.G. is supported in part by grant INTAS-99-1705
and K.S. by INTAS-97-0103.
A.G. thanks J. Ambj{\o}rn for the hospitality at NBI where the part
of the work has been done.

\end{document}